\documentclass[twocolumn,preprintnumbers,amsmath,amssymb]{revtex4}

\pdfoutput=1

\usepackage{todonotes}
\usepackage{SIunits,bm,amsmath}

\usepackage{graphicx}
\usepackage{epstopdf} 
\usepackage{dcolumn}
\usepackage{bm}
\usepackage{amsmath}

\usepackage{color} 

\begin{document}


\title{A novel vortex generator and mode converter  for electrons
}

\author{P. Schattschneider}
\affiliation{Institut f\"ur Festk\"orperphysik, Technische
Universit\"at Wien, A-1040 Wien, Austria}
\email{schattschneider@ifp.tuwien.ac.at }
\affiliation{Ecole Centrale Paris, LMSSMat (CNRS UMR 8579), F-92295 Ch\^atenay-Malabry, France}
\author{ M. St\"oger-Pollach}
\affiliation{University Service Centre for Electron Microscopy,  Technische Universit\"at Wien, A-1040 Wien, Austria}
\author{J. Verbeeck}
\affiliation{EMAT, University of Antwerp, Groenenborgerlaan 171, 2020 Antwerp, Belgium}

\begin{abstract}
A mode converter for electron vortex beams is described. 
Numerical simulations, confirmed by experiment,  show  that  the converter transforms a vortex beam with topological charge $m=\pm 1$ into beams closely resembling Hermite-Gaussian HG$_{10}$ and HG$_{01}$ modes. The converter can be used as a mode discriminator or filter for electron vortex beams.   Combining the converter with a phase plate turns a plane wave into  modes with topological charge $m=\pm 1$.  This combination serves as a  generator of  electron vortex beams of high brilliance.
\end{abstract}


\maketitle
The creation of free electron vortices~\cite{UchidaNature2010,VerbeeckNature2010} opened a new path to the study of matter. The first practical application was  a filter for magnetic transitions, thus facilitating experiments in  energy loss magnetic chiral dichroism (EMCD)\cite{LoydPRL2012,SchattNature2006}. Electron vortices up to topological charge  $m \sim 100$ can now be  produced routinely with holographic masks\cite{McMorranScience2011}, and they may also occur naturally in the  wave function of  electrons after interaction with a crystal~\cite{AllenUM2001}.  Potential applications   range from the study of  chiral structures over manipulation of nanoparticles, clusters and molecules~\cite{XinMuller2010} to spin filters\cite{Karimi2012} and quantum computing~\cite{Groblacher2006}. Their local phase sructure is similar to that of an electron in extremely strong magnetic fields~\cite{AidelsburgerPRL2011} --- of the order of 1000~Tesla for vortices of nm dimension --- which can make them a model system for the study of phenomena at such fields. One of the  unique features of electron vortices is that, owing to their rotational component of probability current they carry   a magnetic moment, even for beams without spin polarization. This, the possibility to focus free electron vortices on a scale of one Angstrom~\cite{VerbeeckAPL2011}, and their strong interaction with matter makes them  attractive.
 
Many properties of free electron vortices can be described by methods applied in singular optics, based on the theory of Nye and Berry~\cite{NyeBerry1974}. 
Optical vortices are now used or proposed in many different research areas such as contrast improvement in astronomy\cite{FooOptLett2005} and microscopy\cite{Furhapter2005},  manipulation of microscopic particles\cite{Friese2001}, control of Bose-Einstein condensates\cite{Andersen2006}, gravitational-wave detection\cite{Granata2010} and quantum  cryptography\cite{Groblacher2006}.  
For a review of optical vortices and their application see\cite{Molina2007,Franke2008}.

For the control  of  vorticity  of electron beams in  transmission electron microscopy (TEM) mode converters 
would be of great interest.  Such   devices change the topological charge of an incident vortex by adding or subtracting charge units. An evident application is mode discrimination: Since vortex modes with charge $m$ and $-m$ have the same spatial intensity distribution, a mode converter adding $\mu$  topological charges would yield modes with charge $\mu+m$ or $\mu-m$, which can be distinguished by their different radial profiles\cite{SchattUm2012a,Karimi2012}. Moreover, a mode converter could increase the vorticity of an electron and thus the magnetic moment, or simply  be used to create a vortex from  an incident plane wave (which has topological charge $m=0$) by adding one unit of charge. The latter aspect is probably the most important one because the whole current would go into one vortex, contrary to the present holographic technique that blocks  half of the intensity in the amplitude mask and distributes the remaining intensity into 3 fundamental modes and several higher harmonics. It goes without saying that  such a device could exploit the entire brilliance of the electron source, resulting in an intensity gain of about an order of magnitude.


In laser optics, mode converters are used to change the topological charge of a beam. The idea is based on the linearity between  Laguerre-Gaussian (LG) modes (which possess  topological charge) on the one hand and Hermite-Gaussian (HG) modes  (which dont) on the other hand, combined with the phase shifting action of cylinder lenses. A particular setting of the lens parameters in  combination with two cylinder lenses converts HG modes into LG modes\cite{BeijersOptComm1993}. 

Since an equivalent setup in the TEM is difficult to realise we propose here a simple scheme for an electron mode converter, transforming electron beams with topological charge $m= \pm 1$ into HG modes and vice versa. A modification  adding   a phase plate  can convert a plane wave into a mode with topological charge $m=\pm 1$, and thus create a vortex.

Under paraxial conditions with an incident wave $\psi_1$ at a distance $z1$ from the front focal plane (FFP) of a lens the wave function in the observation plane that is at a distance $z2$ from the back focal plane (BFP) is given by\cite{Glaser1952}
\begin{equation}
 \psi_2({\bf x})=e^{i k x^2 z_1/2 f^2} \int_\Pi \psi_1({\bf q}) 
  e^{i\chi({\bf q})} e^{i q^2 z_2/2k} e^{-i {\bf q x}} d^2q 
\label{psi2}
\end{equation}
 with the phase 
 $$
 \chi({\bf q})=df (q_x^2-q_y^2)/2k +C_s q^4/4 k^3,
 $$
  and the integral is over the aperture function $\Pi$.
 In order to  distinguish the front focal and back focal planes we use the variable $\bf q$ in the FFP and the variable $\bf x$ in the BFP.
 The parameter $df$ is the astigmatic defocus (the stigmatic axes are in $x$ and $y$ direction), and   $C_s$ is the spherical aberration coefficient of the lens.


When both wave functions are in their respective focal planes Eq.\ref{psi2} collapses to the well known  Fourier transform between object and diffraction:
\begin{equation}
 \psi_2=FT[\psi_1 
 e^{i \chi}].
\label{psi2b}
\end{equation}
Calculations are based on Eq.\ref{psi2} and --- where it applies -- on Eq.\ref{psi2b}.


In the following we adapt the approach of
Beijersbergen {\it et al.}~\cite{BeijersOptComm1993,PadgettJOpt2002} for optical vortices which in the paraxial regime  obey Eqs.~\ref{psi2}, \ref{psi2b}~\cite{BornWolf}. Any Laguerre-Gaussian (LG)  mode can be written as a linear superposition of  Hermite-Gaussian (HG) beams. In the present context we are interested in modes with topological charge $|m|=1$. In the notation of Beijersbergen\cite{BeijersOptComm1993}
\begin{equation}
LG_{10}=\frac{1}{\sqrt 2} (HG_{10}-i HG_{01}) 
\label{LG10}
\end{equation}
\begin{equation}
LG_{01}=\frac{1}{\sqrt 2} (HG_{10}+ i HG_{01}) .
\label{LG01}
\end{equation}
LG$_{nm}$  beams carry topological charge $n-m$. 

The basic mode converter imposes a {\it relative} phase shift of $\pi$ between the two components on the right hand side of Eq.~\ref{LG10}. That  transforms the wave into
\begin{equation}
\frac{1}{\sqrt 2} (HG_{10}-i^2 HG_{01})=\frac{1}{\sqrt 2} (HG_{10}+ HG_{01}).
\label{HG10b}
\end{equation}
Noting that 
\begin{equation}
HG_{01}(\xi,\eta,z)=\frac{1}{\sqrt 2} (HG_{10}(x,y,z)+HG_{01}(x,y,z))
\label{HG11}
\end{equation}
where new rescaled axes $\xi, \eta$, rotated by 45 degrees with respect to $x,y$ are used:
$\xi=(x+y)/\sqrt 2 \, , \qquad
\eta=(x-y))/\sqrt 2
$,
the output of the converter, Eq.~\ref{HG10b} is  a  $HG_{01}$ mode with its axis along the 45 degree direction. The same phase shift applied to  Eq.~\ref{LG01} results in an HG$_{10}$ mode, again with its axis along the 45 degree direction. So, the action of the phase shifting converter can be expressed in short as
 $$
 LG_{10} \overset{\pi}{\rightarrow} HG_{01} \qquad
  LG_{01} \overset{\pi}{\rightarrow} HG_{10} .
$$
 
Applying this idea to electrons raises two questions: 1) Are the vortex beams that can now be produced with holographic masks or emerging from a specimen after spin-polarized transitions~\cite{SchattPRB2012} a sufficiently precise approximation to the $LG_{10}$ and $LG_{01}$ modes? 2) How can we impose a phase shift of $\pi/2$ between the  components?

The vortex beams emerging from the holographic masks are superpositions of Bessel beams.  A comparison of their radial distribution with an LG$_{10}$ mode in the BFP shows that they are indeed very similar when the correct beam waist is chosen.  
The required phase shift between modes (task 2 above) is achieved exploiting the Gouy phase\cite{Gouy1890} of astigmatic beams. For HG beams of order $(nm)$ it is 
\cite{BeijersOptComm1993}
 \begin{equation}
\phi=(n+1/2) \arctan(\frac{z-z_x}{z_R})+(m+1/2)\arctan(\frac{z-z_y} {z_R}) 
\label{Gouynm}
\end{equation}
where $z_x, z_y$ are the positions of the astigmatic line foci.

\begin{figure}
	\centering	\includegraphics[width=\columnwidth]{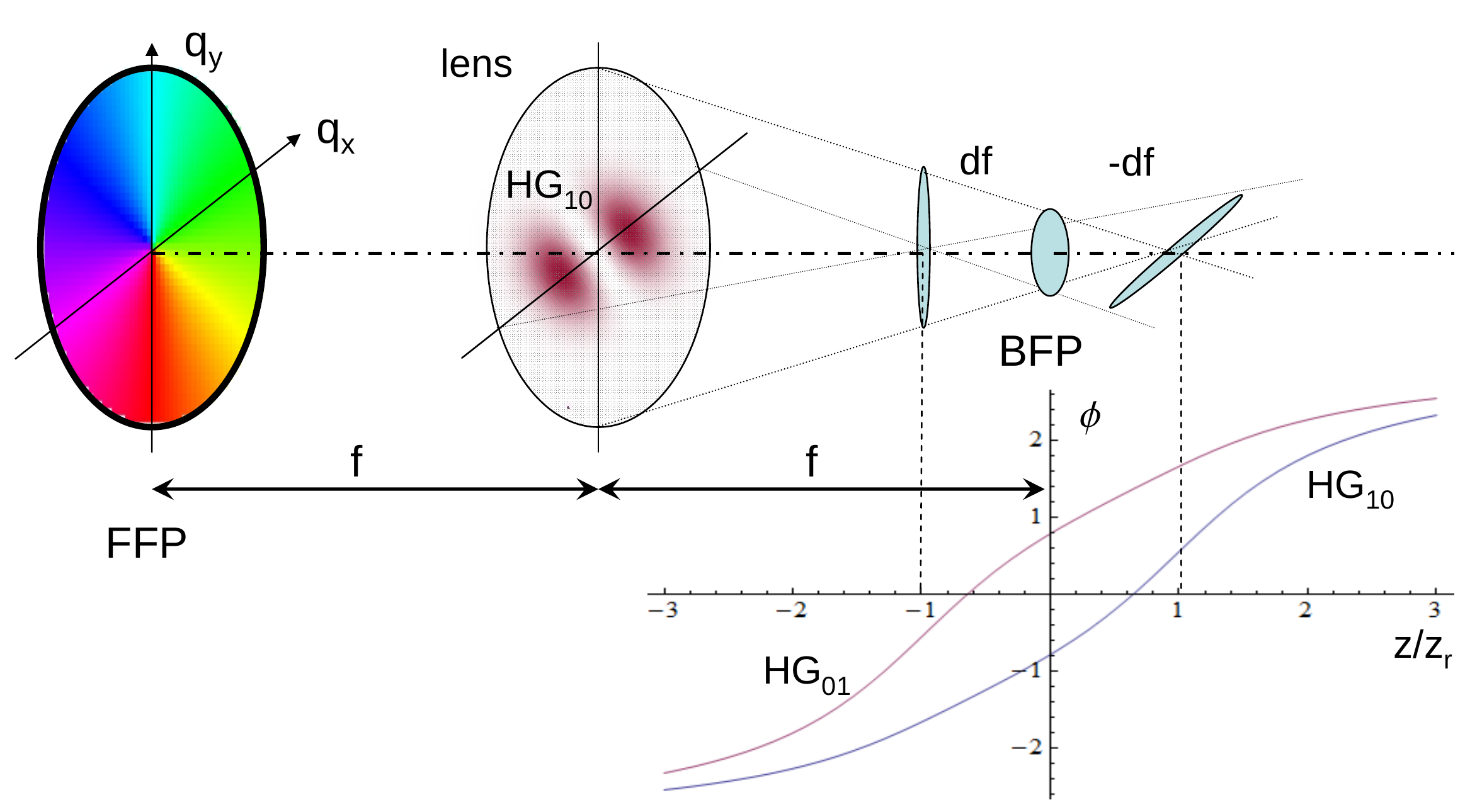}
	\caption{Electron optical arrangement: A vortex fills the aperture in the FFP of a magnetic astigmatic lens. The phase is color coded (rainbow wheel). The astigmatic line foci along the $x$ and $y$ axes  are at a mutual distance of $2 df$. The vortex is composed of two phase shifted HG-like components - Eq.~\ref{LG10} -  one of which is symbolised as it passes the lens.  Also drawn are the Gouy phases for the HG$_{01}$ and the HG$_{10}$ components for the case that the astigmatism equals the Rayleigh range, $df=z_R$. The observation plane is in the BFP midway between the line foci; there, the phase difference between the two  HG modes is $\pi/2$.}
	\label{fig:Fig1}
\end{figure}

Putting the beam waists  at $z_x=z_R$, $z_y=-z_R$ as in Fig.\ref{fig:Fig1} the {\it relative} Gouy phase shift between the fundamental modes HG$_{10}$ and  HG$_{01}$ is, according to Eq.\ref{Gouynm} 
\begin{equation}
\Delta \phi= \arctan(\frac{z-z_R}{z_R})-\arctan(\frac{z+z_R}{z_R})
\label{GouyDelta}
\end{equation}
which at $z=0$ is $\pi/2$. In other words, placing the observation plane at $z=0$ in a lens with astigmatism $df=z_R$ the phase shift between the two components is  $\pi/2$. 
A  complication is that the Gouy phase for non-Gaussian beams does not follow Eq.~\ref{GouyDelta}.
Comparison of numerically obtained Gouy phases with that of a Gaussian beam of the same FWHM as the Airy disk results in an optimized astigmatism of $df=220$~nm, almost identical with the Rayleigh range of the Gaussian beam.
An astigmatic defocus of that value should  induce the correct phase shift between the $x$ and $y$ components of the focused plane wave.

\begin{figure}
	\centering
\includegraphics[width=0.29\columnwidth]{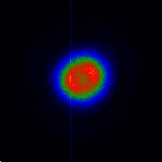}
\includegraphics[width=0.29\columnwidth]{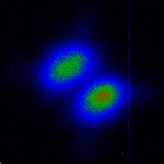}	\includegraphics[width=0.3\columnwidth]{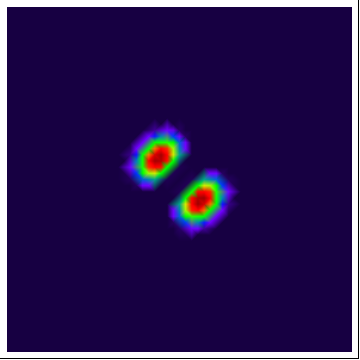}
\includegraphics[width=0.29\columnwidth]{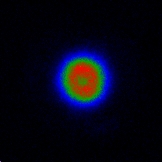}
\includegraphics[width=0.29\columnwidth]{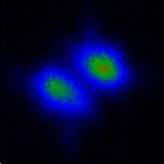}	\includegraphics[width=0.3\columnwidth]{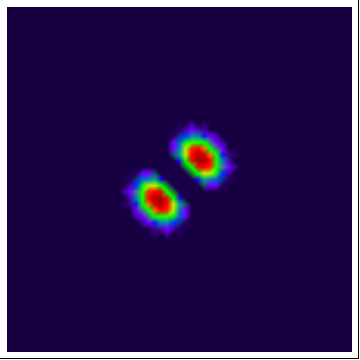}			
	\caption{Experimental mode conversion. Upper row: Vortex with topological charge $m=-1$ after passage of a tunable lens. Left:  Astigmatism $df= 0$, yielding a focused vortex. Middle:  Astigmatism $df$=220~nm. Right: Simulation for 200~kV. Bottom row: Same for $m=1$. Squares have a side length of 5~nm. Intensities in all Figs. in false colors.}
	\label{fig:Fig3}
\end{figure}

\begin{figure}
	\centering
\includegraphics[width=0.29\columnwidth]{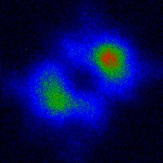}		\includegraphics[width=0.3\columnwidth]{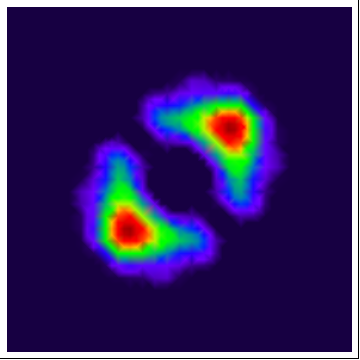}			
	\caption{Vortex with topological charge $m=1$ after passage of a tunable lens with astigmatism $df= 700$~nm.  Left: Experiment; Right: Simulation for 200~kV. Scale as in Fig.~\ref{fig:Fig3}. }
	\label{fig:Fig4}
\end{figure}

\begin{figure}
	\centering	\includegraphics[width=\columnwidth]{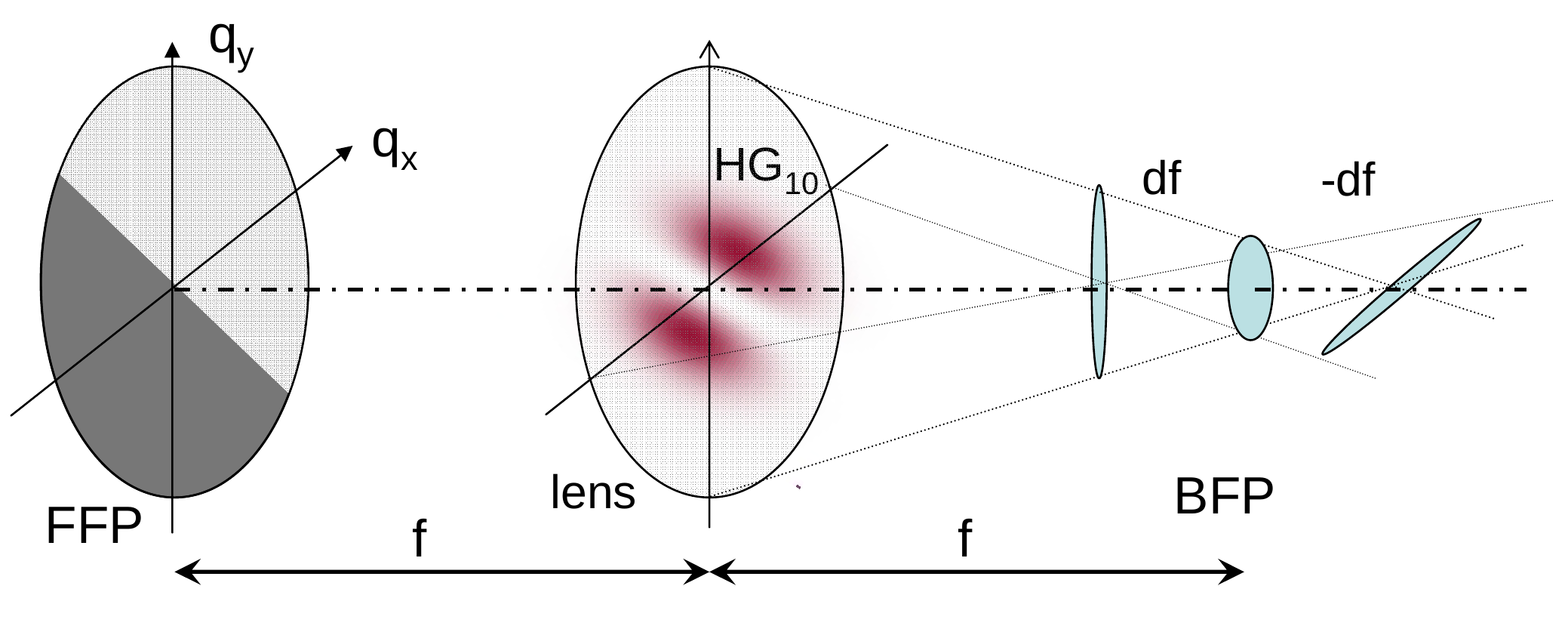}
\includegraphics[width=0.30\columnwidth, angle=270]	{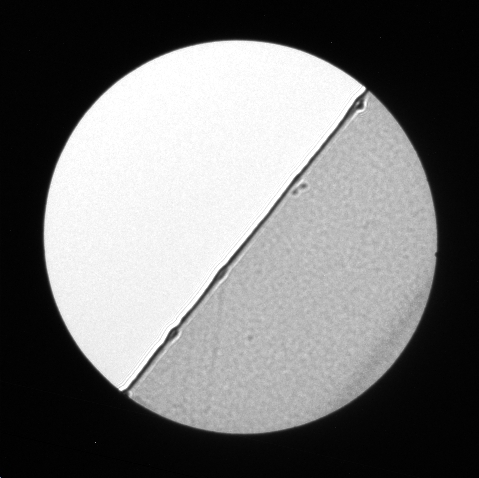}	\includegraphics[width=0.30\columnwidth, angle=270]{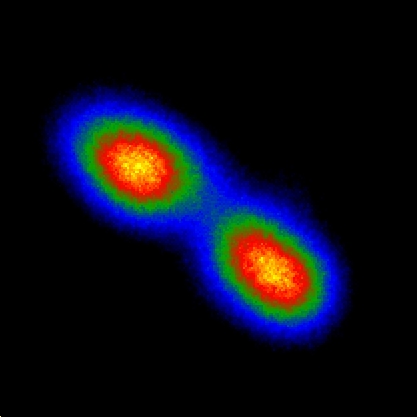}
	\caption{Geometry for a vortex generator. A phase plate in the FFP imposes a phase shift of $\pi$ on one half of the incident plane wave. This makes a rotated HG$_{10}$-like beam, here symbolised as it passes the lens. The Gouy phase shift of the astigmatic lens creates an LG$_{10}$ beam in the BFP. 
	b) Shadow image of the phase plate. c) HG$_{10}$-like beam in the BFP obtained with the phase plate and no astigmatism.}
	\label{fig:Fig1c}
\end{figure}

Experiments were performed on a TECNAI F20 microscope ($C_s=1.2$~mm) at 200~kV (Figs.~\ref{fig:Fig3},~\ref{fig:Fig4}) and at 86~kV (Figs.~\ref{fig:Fig1c},~\ref{fig:experiment}). The input vortices were produced with a holographic fork mask in the FFP of the condenser lens. When the lens has no astigmatism the well-known focused vortices are found in the BFP. The left column of Fig.~\ref{fig:Fig3} shows these beams. Tuning the astigmatism to $df=z_R$, the vortices with topological charge $m=\pm 1$ are converted into HG$_{10}$ -like modes rotated by $\pm \pi/4$ about the optic axis, as shown in the center panels.  The agreement with simulation (right panels)  is excellent. Deviations (lagely blurring) are caused by remaining lens aberrations, incomplete coherence, and the missing mode matching in this geometry~\footnote{In the observation plane the radii of curvature of the constituent waves have different sign. Mode matching uses two cylinder lenses to avoid this problem.}.
We found that the structure of the converted modes is surprisingly stable under variations of astigmatism. Fig.~\ref{fig:Fig4} shows the converted $m=1$ mode for a larger astigmatism of $df=700$~nm. The broken azimuthal symmetry in the HG modes seen in Fig.~\ref{fig:Fig3} can be used to analyse the topological charge on a sub-nm scale, with possible applications in crystallography~\cite{AllenUM2001}, chirality~\cite{XinMuller2010} and spin polarized electronic  transitions~\cite{LoydPRL2012,SchattPRB2012}.

Since such converters operate also in "reverse" mode, transforming HG into LG beams, one can use the device as a vortex generator without the need of an amplitude mask that blocks half of the intensity. The entire  signal  would go into one vortex, rather than into 3 fundamental modes and higher harmonics. The difficulty lies in the experimental realisation of HG electron beams. Contrary to laser optics where Gaussian beam profiles occur quite naturally, in electron optics there are plane or convergent waves limited by round apertures. But since the salient feature of the HG$_{nm}$ modes is the phase shift of $\pi$ between lobes one can hope for a reasonable result using a phase plate.  Such devices,  proposed 1942 in order to increase contrast by inducing a phase shift of $\pi/2$ \cite{Zernike1942b} have seen a revival in the study of biological specimens\cite{Nagayama2008}. Here, we use a Hilbert plate\cite{Nagayama2008,Danev2002}  (which  induces a phase shift of $\pi$ between the two lobes of a beam) not for contrast improvement but for phase inversion. Ideally that beam should resemble a $\pi/4$ rotated HG$_{10}$ beam in the BFP. Fig.~\ref{fig:Fig1c}c is an experimental image obtained with a Hilbert plate. The two lobes of the focused beam are clearly discernible. The corresponding wave functions reveal a phase difference of $\pi$.  According to Eq.~\ref{HG11} the wave function is proportional to a superposition $HG_{10} + HG_{01}$. Fig.~\ref{fig:Fig1c}b is an electron microscopical shadow image  of the phase plate, cut from a commercially available Si-nitride thin film, projected on the probe forming aperture in the FFP.   

When the vortex generator is activated the Gouy phase shift of the astigmatic lens converts this HG mode into an LG$_{10}$ beam in the BFP. 
 \begin{figure}
	\centering	\includegraphics[width=0.49\columnwidth]{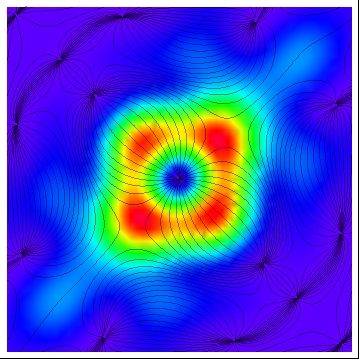}	\includegraphics[width=0.49\columnwidth]{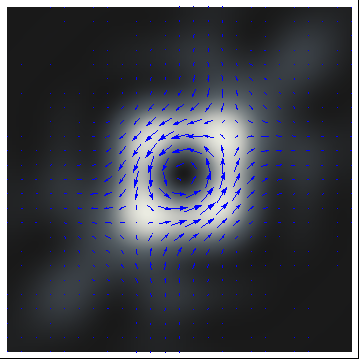}	
	\caption{Ideal output of the vortex generator. 
	Left: Intensity and cophasal lines increasing from 0 to 2$\pi$ with the phase singularity in the center, characteristic for a topological charge $m=1$. Right: Same with superimposed ring shaped quantum mechanical current density. Scale as in Fig.~\ref{fig:Fig3}. }
	\label{fig:v0}
\end{figure}
 \begin{figure}
	\centering	\includegraphics[width=0.37\columnwidth]{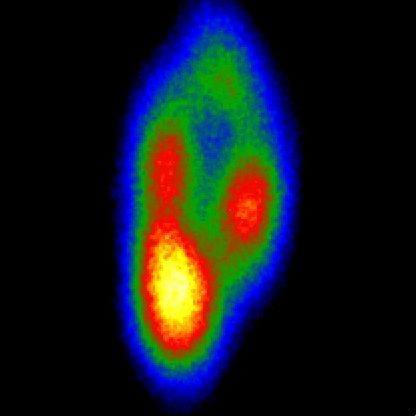}	\includegraphics[width=0.40\columnwidth]{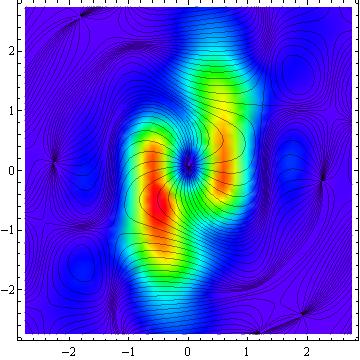}	
	\caption{Left: Experimental vortex generator with nominal $df$ as in Fig.~\ref{fig:v0}. Right: Optimized simulation (20\% absorption, $df=500$~nm, defocus 400~nm). Despite astigmatic distortions the central phase singularity is still visible. }
	\label{fig:experiment}
\end{figure}
Simulations based on Eq.\ref{psi2} for an ideal phase plate
are shown in Fig.~\ref{fig:v0}.
The phase singularity in the center of the left panel where all cophasal lines merge proves the presence of topological charge ($m=1$ in the figure). The missing mode matching possibility in this geometry and the deviation from the Gaussian profiles cause the slight anisotropy of the charge density. However the quantum mechanical current density $j=\rho \nabla \varphi$ (left panel) demonstrates clearly the vortex character of the output beam.  

Results of  a demonstration experiment are shown in Fig.~\ref{fig:experiment}a for a nominal astigmatism  corresponding to Fig.~\ref{fig:v0}. A strong astigmatic defocus is present in vertical direction, nevertheless  4 local maxima corresponding to the  4-fold symmetry predicted in Fig.~\ref{fig:v0} can be seen. After variation of the simulation parameters (Fig.~\ref{fig:experiment}b) it turned out that the main reasons for the disagreement are the strong absorption in the phase plate and a remaining defocus that is probably caused by crosstalk of the magnetic field of the objective lens with that of the condenser. 

Optimizing the vortex generator will be challenging. A homogeneous phase plate of the necessary size, stability and phase shift is difficult to produce and to maintain (beam damage and contamination will deteriorate the thin transparent film rapidly,) but is not out of reach. Lack of mode matching, the difficulty to align the astigmatic axes with the edge of the phase plate, and the unavoidable cross-talk of the objective lens field into the condenser limit the performance. A combination of two astigmatic lenses will probably  improve the quality of the vortex generator. 

In summary we have demonstrated   mode conversion for vortex electrons. The method can be used for  discrimination of topological charge or for mode filtering. A variant including a phase plate  can potentially  generate vortex beams from incident plane electron waves with intensities surpassing that of the  established fork mask technique by  an order of magnitude. 


{\bf Acknowledgements:} 
P.S. acknowledges financial suppport of the Austrian Science Fund, project I543-N20. J.V. acknowledges support from the European Research Council under the 7th Framework Program (FP7), ERC grant Nr.~246791 - COUNTATOMS and ERC Starting Grant 278510 - VORTEX. 

%

\end{document}